\documentclass[5p,sort&compress]{elsarticle}

\usepackage[usenames]{color}
\usepackage[hidelinks]{hyperref}

\journal{Phys.\ Lett.\ B}

\usepackage{amsmath,amssymb}

\usepackage[]{cleveref}
\Crefname{equation}{Eq.}{Eqs.}
\Crefname{figure}{Fig.}{Figs.}
\crefformat{plural}{#2eqs.~(#1)#3}
\crefname{section}{Sect.}{Sects.}

\def\be#1\ee{\begin{align}#1\end{align}}

\newcommand{\ie}{i.e.}

\renewcommand{\ge}{\geqslant}

\DeclareMathOperator\Ima{Im}
\newcommand{\AdS}{\text{AdS}}

\begin{document}

\begin{frontmatter}

\title{Van der Waals-like Behaviour of Charged Black Holes\\
and Hysteresis in the Dual QFTs}

\author{Mariano Cadoni, Edgardo Franzin, and Matteo Tuveri}
\address{Dipartimento di Fisica, Universit\`a di Cagliari
\& INFN, Sezione di Cagliari\\
Cittadella Universitaria, 09042 Monserrato, Italy}

\begin{abstract}
Using the rules of the AdS/CFT correspondence, we compute the spherical
analogue of the shear viscosity, defined in terms of the retarded Green
function for the stress-energy tensor for QFTs dual to five-dimensional
charged black holes of general relativity with a negative cosmological
constant. We show that the ratio between this quantity and the entropy
density, $\tilde\eta/s$, exhibits a temperature-dependent hysteresis.
We argue that this hysteretic behaviour can be explained by the Van der
Waals-like character of charged black holes, considered as thermodynamical
systems. Under the critical charge,
hysteresis emerges owing to the presence of two stable states (small and large
black holes) connected by a meta-stable region (intermediate black holes).
A potential barrier prevents the equilibrium path between the two stable
states; the system evolution must occur through the meta-stable region, and a
path-dependence of $\tilde\eta/s$ is generated.
\end{abstract}

\end{frontmatter}

The investigation of black brane configurations with holographic duals has become
an important source of information both for the understanding of fundamental
features of the gravitational interaction and for the description of strongly
coupled QFTs~\cite{Kovtun:2003wp,Kovtun:2004de,Policastro:2002se,%
Hartnoll:2008vx,Horowitz:2010gk}.
In this context, the hydrodynamic limit of holographic QFTs plays a very
important role, because it allows computing transport coefficients, like the
shear viscosity to entropy density ratio $\eta/s$, in the strongly
coupled regime of the QFT\@.
This has led to the proposal of a fundamental bound $\eta/s\ge1/4\pi$, known as
the KSS bound~\cite{Kovtun:2003wp}, which found support both from string
theory~\cite{Kovtun:2004de} and quark-gluon plasma experimental
data~\cite{Song:2010mg}.
By now, it is well-known that the KSS bound is violated by higher curvature
terms in the Einstein-Hilbert action~\cite{Brigante:2007nu} or by breaking of
translational or rotational symmetry of the black brane
background~\cite{Erdmenger:2010xm,Rebhan:2011vd,Mamo:2012sy,Davison:2015taa,%
Hartnoll:2016tri,Alberte:2016xja,Burikham:2016roo,Cadoni:2016hhd,Liu:2016njg}.
Typically, when the KSS bound is violated, $\eta/s$ exhibits a non-trivial
dependence on the temperature~\cite{Cremonini:2012ny}.

Until now, these investigations have been restricted to planar topologies in
the bulk (black branes) and have not concerned spherical topologies (black
holes). The main obstruction to this generalisation is the absence of the usual
hydrodynamic limit for QFTs dual to spherical black holes.
Indeed, differently from the black brane case, the spherical geometry of the
horizon breaks the translational symmetry in the dual QFT
preventing the existence of conserved charges.
However, it is still possible to define a relativistic hydrodynamics in curved
spacetimes without translational symmetry as an expansion in the derivatives of
the hydrodynamic fields of the stress-energy tensor~\cite{Baier:2007ix} and
a related Kubo formula for the shear viscosity.

The hydrodynamic limit of a QFT living on a curved spacetime can be defined
in the same way as for a QFT in the plane.
We just consider the system at large relaxation times (small frequencies) and
large scales compared to the microscopic scale of the system.
When the latter is unknown, we can still give a thermal description of the
system and associate this microscopic scale with the inverse of the temperature
$T$. Thus, the hydrodynamic limit corresponds to consider excitations of the
system with wavelength $\lambda\gg1/T$. In this limit, the macroscopic behaviour
of the QFT living in a curved background is described by a stress-energy tensor,
which can be written as~\cite{Baier:2007ix,Romatschke:2009kr}
\be\label{dissipativeT}
T^{ab}=\left(\epsilon + P\right) u^au^b + P g^{ab} + \Pi^{ab}\,,
\ee%
where $\epsilon$ and $P$ are the energy density and the thermodynamical pressure
and $u^a$ is the fluid velocity, usually considered in the frame in which the
fluid is at rest. The tensor $\Pi^{ab}$ contains all the dissipative contributions
to the stress-energy tensor. At first order in the velocity expansion, it depends
on the transport coefficient $\kappa$, the relaxation time~$\tau_\Pi$ and the
shear viscosity~$\eta$.

The previous considerations hold for a QFT in a generic curved space.
Working in the AdS/CFT framework, we can apply \cref{dissipativeT} to a four
dimensional CFT dual to a five dimensional AdS bulk spherical black
hole~\cite{Gubser:1998bc,Witten:1998qj,Cho:2002hq,Neupane:2009zz}.
To derive a Kubo formula for CFTs living on the boundary of $\AdS_5$, whose
spatial section is the three-sphere, we consider small perturbations around the
boundary background metric, \ie\ $g_{ab}=\bar{g}_{ab} + h_{ab}$.
Without loss of generality, we use transverse and traceless perturbations, which
in turn, depend on time and the angular directions. Under these assumptions and
in linear approximation, \cref{dissipativeT} becomes
\be\label{shearmodes}
T^{ij}=-Ph_{ij}- \eta\dot{h}_{ij} + \eta\tau_{\Pi}\ddot{h}_{ij}
-\frac{\kappa}{2}\left[\ddot{h}_{ij} + L^2\Delta_L h_{ij}\right],
\ee%
where $\Delta_L$ is the Lichnerowicz operator and $L$ is the $\AdS_5$ length.
We choose a harmonic time dependence for the perturbation and we expand it
in hyperspherical harmonics.
We now extract the retarded Green function for the spatial components of the
stress-energy tensor $T^{ij}$ in the tensor channel and, from \cref{shearmodes},
we read
\be\label{GreenFunction}
G_{T^{ij}T^{ij}}^R(\omega,\ell) = -P -i\omega\eta - \omega^2\eta\tau_{\Pi}
-\frac{\kappa}{2}\left(\omega^2+L^2\gamma\right),
\ee%
where $\gamma\equiv\ell(\ell+2)-2$ is the eigenvalue of the Lichnerowicz
operator and $\ell=1,2,3,\ldots$ is the first number associated with the
hyperspherical harmonic expansion. \Cref{GreenFunction} allows us to derive a
Kubo formula for the analogue of the shear viscosity $\tilde\eta$ for a
relativistic QFT on a spatial spherical background as
\be\label{analogueKubo}
\tilde\eta = -\lim_{\omega\to0} \frac{1}{\omega}
\Ima G_{T^{ij}T^{ij}}^R(\omega,\ell\to\ell_0)\,,
\ee%
where $\ell_0$ is the smallest eigenvalue of the Lichnerowicz operator and
$\omega$ is the frequency of the perturbation. Notice that the only difference
of \cref{analogueKubo} with the planar case is the evaluation of the retarded
Green function in $\ell\to\ell_0$ instead of wavenumber $k\to0$.

It is important to stress that, with respect to the planar case, we have an
additional contribution to the stress-energy tensor~\eqref{shearmodes}. This is
rather expected in view of the breaking of translational invariance. However,
this $\kappa$-term does not contribute to the shear viscosity.

Let us now compute the spherical analogue viscosity to entropy density ratio
$\tilde\eta/s$ for the QFT dual to a five-dimensional spherically symmetric
charged black hole of general relativity with a negative cosmological constant.
Differently from black branes, black holes have a rich thermodynamical phase
structure, characterised by different stable or metastable phases (small and
large black holes, thermal AdS). As a consequence, one naturally expects the
correlators~\eqref{GreenFunction} and even more the ratio $\tilde\eta/s$
to keep track of this rich phase structure.
We find a hysteretic behaviour of $\tilde\eta/s$ as a function of the temperature
and we explain it in terms of the Van der Waals-like behaviour of this class of
black holes when considered as thermodynamical systems~\cite{Chamblin:1999tk,%
Chamblin:1999hg}.
Detailed calculations and generalisation to Gauss-Bonnet neutral and charged
black holes are presented in Ref.~\cite{Cadoni:2017ktd}.

The line element of the five-dimensional anti de Sitter-Reissner-Nordstr\"om
(AdS-RN) black hole is
\be%
ds^2 &= g_{ab}^{(0)}dx^a dx^b =
-f(r)\,dt^2+\frac{dr^2}{f(r)} + r^2\,d\Omega_3^2\,,\label{ds}\\
f(r) &= 1 + \frac{r^2}{L^2} - \frac{8 M}{3\pi r^2}
+ \frac{4\pi Q^2}{3r^4}\,,\label{fRN}
\ee%
where $d\Omega_3^2$ is the line element of the 3-sphere, $L$ is the AdS length,
$M$ is the black hole mass and $Q$ its charge.

The Van der Waals-like liquid/gas phase transition for AdS-RN black holes can
be understood by discussing the black hole free energy or by considering the
relation between the temperature and the radius of the black hole
\be\label{temperature}
T(r_+) = \frac{r_+}{\pi L^2} + \frac{1}{2\pi r_+} - \frac{2 Q^2}{3 r_+ ^5}\,.
\ee%

\begin{figure}[!ht]
\centering
\includegraphics[width=0.42\textwidth]{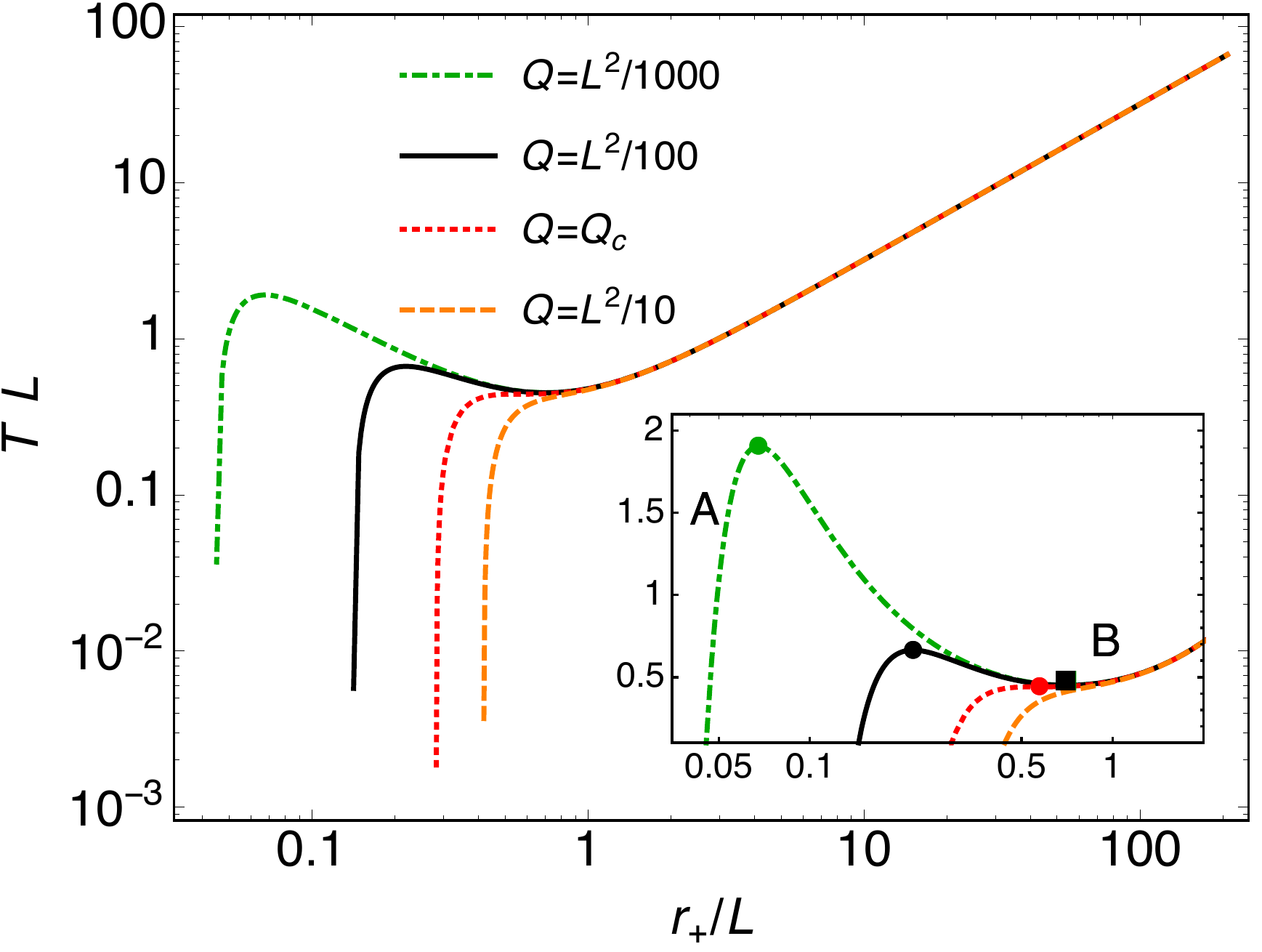}
\caption{Plot of the function $T(r_+)$ for selected values of $Q$ above, at and
below the critical charge $Q_c$. Inset: Zoom in the region where the function
has local extrema. The dots and squares mark the critical temperatures. A and B
denote, respectively, the small and large black hole stable regions. Notice
that the minima of the $Q=L^2/1000$ and $Q=L^2/100$ curves almost coincide.}
\label{fig:Tr}
\end{figure}

As the charge of the black hole decreases to the critical charge
$Q_c=L^2/6\sqrt{5\pi}$, the black hole undergoes a second-order phase transition.
Below $Q_c$ the system is characterised by the presence of two stable
states (small and large black holes) connected through a meta-stable region of
intermediate black holes --- see \cref{fig:Tr}. The phase transition small/large
black holes is a first-order one~\cite{Chamblin:1999tk, Chamblin:1999hg}.

Following the rules of the AdS/CFT correspondence, to calculate $\tilde\eta$ for
the QFT dual to the five-dimensional AdS-RN black hole we consider transverse
and traceless perturbations about the background metric~\eqref{ds},
$g_{ab}=g^{(0)}_{ab} + h_{ab}$ with $h_{ab}=0$ unless $(a,b)=(i,j)$ and
$h_{ij}(r,t,x)=r^2\,\phi(r,t)\,\bar{h}_{ij}(x)$, being $\bar{h}_{ij}$ the
eigentensor of the Laplacian operator built on the $3$-sphere. Such perturbations
are gauge-invariant and by linearising the Einstein field equations, the angular
part decouples~\cite{Gibbons:2002pq,Dotti:2004sh,Dotti:2005sq}. By assuming a
harmonic time dependence for the perturbation, $\phi(r,t)=\psi(r)\,e^{-i\omega t}$,
one finds the linear second-order differential equation for $\psi(r)$
\be\label{eom}
\frac{1}{r^3}\frac{d}{dr}\left[r^3 f(r)\,\frac{d\psi(r)}{dr}\right]
+\left[\frac{\omega^2}{f(r)}-m^2(r)\right]\psi(r)=0\,,
\ee%
where $f(r)$ is given by \cref{fRN}, the mass term for the perturbation is
$m^2(r)={[4-\ell(\ell+2)]}/{r^2}$ and $\ell$ are the eigenvalues of the Laplace
operator on the 3-sphere.
The presence of a non-vanishing mass term in \cref{eom} is a consequence of the
breaking of translational symmetry due to the spherical geometry of the horizon.
In the black brane context, this term is responsible for the violation of the
KSS bound~\cite{Hartnoll:2016tri} and generates a dependence of the shear
viscosity to entropy ratio on the temperature.
\Cref{eom} admits as solutions a non-normalisable and a normalisable mode
that behave asymptotically as $\psi_0\approx 1,\,\psi_1\approx 1/r^4$.

The retarded Green function in \cref{analogueKubo} can be  found  using the
method proposed in Refs.~\cite{Lucas:2015vna,Hartnoll:2016tri} which gives  
a very simple and elegant way for computing correlators in a QFT dual to a
gravitational bulk theory. The spherical analogue viscosity to density
entropy ratio is determined by the non-normalisable mode $\psi_0(r)$
evaluated at the horizon
\be\label{etatos}
\frac{\tilde\eta}{s}= \frac{1}{4\pi}\,\psi_0(r_+)^2\,.
\ee%

\begin{figure}[!ht]
\centering
\includegraphics[width=0.4\textwidth]{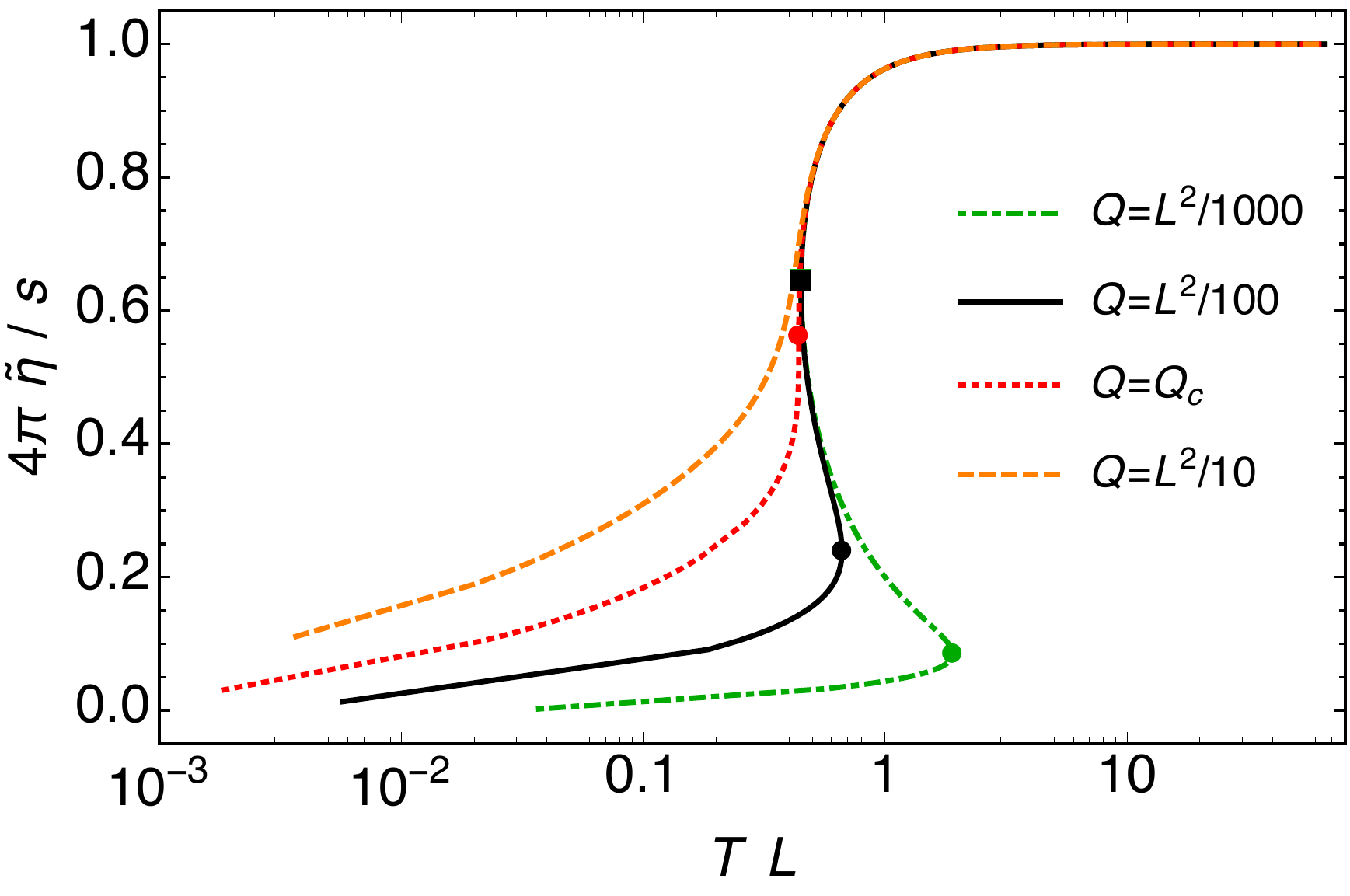}
\caption{Behaviour of $\tilde\eta/s$ as a function of the temperature for dual QFTs
of AdS-RN black holes. We plot $\tilde\eta/s$ for four selected values of the black
hole charge: above, at and below the critical value.
The dots and squares mark the critical temperatures relative to the small/large
black hole first-order phase transition. We have considered
the smallest eigenvalue of the Laplacian, \ie\ $\ell=\ell_0=1$.}\label{fig:GR}
\end{figure}

To determine $\psi_0(r_+)$, we numerically integrate \cref{eom} with $\omega=0$
(supplied by regularity boundary conditions at the horizon) outwards from the
horizon to infinity and then we use a shooting method requiring that
$\psi_0(\infty)=1$. Finally, we compute $\tilde\eta/s$ as a function of $T$
using \cref{temperature,etatos}.
In \cref{fig:GR} we plot our results for $\tilde\eta/s$ for selected values
of the charge $Q$ and we observe that, for \mbox{$Q<Q_c$}, it exhibits a
temperature-dependent hysteresis, after that the second-order Van der Waals-like
phase transition occurs.

In the hydrodynamic, holographic context, a hysteretic behaviour in the shear
viscosity has been already observed for AdS black branes with broken rotational
symmetry and with a p-wave holographic superfluid dual~\cite{Erdmenger:2010xm}.
Moreover, it is known that real fluids may exhibit hysteresis in the $\eta$-$T$
plane, this is, for instance, the case of nanofluids~\cite{Nguyen:2008}.

A quite general thermodynamical explanation of hysteresis associated with phase
transitions has been given and is related to
meta-stabilities~\cite{Knittel:1977,Bertotti:1998}.
Whenever we have at least two stable states --- say $A$ and $B$, respectively on
the left of the maximum and on the right of the minimum of \cref{fig:Tr} ---
connected by a meta-stable region, a potential barrier prevents the evolution of
the system from occurring as an equilibrium path between the two stable states.
Evolution must take place through the meta-stable region and the path $A\to B$
goes from the maximum directly to the state $B$, whereas the path $B\to A$ goes
directly from the minimum to the state $A$ (see the inset in \cref{fig:Tr}).
This is exactly what happens to $\tilde\eta/s$ in the case under consideration.
The state $A$ corresponds to small black holes and the state $B$ to large black
holes, while the meta-stable region (generated by the first-order phase
transition) to intermediate black holes. With this general mechanism, hysteresis
\ie\ path-dependence of $\tilde\eta/s$ is generated.

We conclude this letter with some comments about the relation between our
spherical analogue viscosity $\tilde\eta$ and the usual hydrodynamic viscosity
for QFTs in the plane.
By definition, the shear viscosity is a transport coefficient that measures the
momentum diffusivity due to a strain in a fluid.
Its definition is strictly related to the translational symmetries of the
system which lead to the conservation of momentum and to an associated conserved
current from which one can derive the Fick's law of diffusion~\cite{Son:2007vk}.
This is no longer
true for systems that break translational invariance where the hydrodynamic
interpretation in terms of conserved quantities falls. However, as shown in
Ref.~\cite{Baier:2007ix}, hydrodynamics can be defined as an expansion in the
derivatives of hydrodynamic fields (like the fluid velocity).
This allows one to define the shear viscosity through a Kubo formula also for
QFTs on a spatially curved background, where the stress-energy tensor is only
\emph{covariantly} conserved.
There is an additional conceptual difficulty in defining the hydrodynamic limit
of a QFT on the sphere due to the compactness of the space.
In fact, in a compact space, the usual hydrodynamic limit as an effective theory
describing the long-wavelength modes of the QFT has not a straightforward
interpretation. Our proposal is that for QFTs living on a sphere dual to bulk
black holes, the hydrodynamical, long wavelength modes can be described by the
$\ell\to\ell_0$ modes that probe large angles on the sphere. This is in analogy
with the $k\to0$ modes for a QFT dual to bulk black branes which probe large
scales on the plane.

\bigskip%
We thank Luciano Colombo, Riccardo Dettori and Piero Olla for valuable
discussions and comments.

\end{document}